# Field-effect mobility enhanced by tuning the Fermi level into the band gap of $Bi_2Se_3$

Peng Wei, Zhiyong Wang, Xinfei Liu, Vivek Aji, and Jing Shi

Department of Physics & Astronomy, University of California, Riverside, CA 92521

By eliminating normal fabrication processes, we preserve the bulk insulating state of calcium-doped $Bi_2Se_3$ single crystals in suspended nanodevices, as indicated by the activated temperature dependence of the resistivity at low temperatures. We perform low-energy electron beam irradiation (<16 keV) and electrostatic gating to control the carrier density and therefore the Fermi level position in the nanodevices. In slightly p-doped $Bi_{2-x}Ca_xSe_3$ devices, continuous tuning of the Fermi level from the bulk valence band to the band-gap reveals dramatic enhancement (> a factor of 10) in the field-effect mobility, which suggests suppressed backscattering expected for the Dirac fermion surface states in the gap of topological insulators.

Bi$_2$Se$_3$ stands out as an excellent prototypical material among 3D topological insulators (TI), for it has only a single Dirac cone bridging a large band gap (~ 0.3 eV at 0 K)[1]. Although its massless Dirac fermion surface states have been confirmed by surface-sensitive spectroscopic techniques[2-6]; however, the distinct transport properties of the surface states have not yet been clearly demonstrated. Owing to the overwhelmingly high density of the bulk states that coexist with the surface states, undoped Bi$_2$Se$_3$ behaves as a *de facto* metal, with its Fermi level ($E_F$) located in the bulk conduction band. Calcium (Ca) or antimony (Sb) doping[7-11] can lower $E_F$ into the band gap of bulk Bi$_2$Se$_3$ crystals as evidenced by the thermally activated electrical conductivity. Recently, much progress has been made in TI nanodevices[12-21]. However, strong metallic resistivity, i.e. d$\rho$/dT>0, is found in most nanofabricated devices.

Although the mechanism for observed excess electrons in Bi$_2$Se$_3$ is not yet completely understood, it is believed to be caused by the tendency of Se vacancy formation. In nanofabrication, heating, electron beam writing and other processes may promote Se vacancy formation, which leads to even a higher density of electrons and consequently gives arise to a stronger metallic behavior in resulting nanodevices. To avoid this problem, we bypass the regular nanofabrication processes and successfully maintain the bulk insulating state achieved by Ca-doping in the starting material. To establish a mobility reference, we deliberately choose a p-type metallic state as the initial state, which should have the mobility of bulk carriers. We then control the hole density and therefore the position of $E_F$ using both e-beam irradiation (EBI) and electrostatic gating. We are able to systematically track the mobility change as the nanodevice transforms itself from a metal to an insulator, i.e. from the bulk carrier to surface carrier dominated transport. Additionally, since the Dirac point of the surface states in Bi$_2$Se$_3$ is much closer to the valence band than the conduction band[1-4], it is easier to access the Dirac point from the valence band side.

We first grow high-quality Ca-doped Bi$_2$Se$_3$ single crystals, i.e. Bi$_{2-x}$Ca$_x$Se$_3$, by a multistep heating method described previously[7]. Both the resulting carrier type and carrier density can be tuned by varying the Ca-doping level. At x=0.012, the optimal doping level, a bulk insulating state is obtained as characterized by the highest resistivity (~100 m$\Omega \cdot$cm at ~20 K), a negative temperature coefficient of the resistivity, and the lowest inter-band infrared absorption energy. In order to preserve the insulating state of the starting bulk material, we place exfoliated $Bi_{2-x}Ca_xSe_3$ flakes onto pre-patterned electrodes to form devices without requiring any heating or uncontrolled e-beam exposure. We fabricate two types of pre-patterned electrodes (Fig. 1a) using regular lithography: one with four 0.7 $\mu$m-wide Ti (10 nm)/Au (40 nm) parallel lines and separated 1 $\mu$m from each other; the other with two orthogonal lines with 3$\mu$m long gap on each at the crossing. The first type is suitable for four-terminal resistivity measurements while the second type for Hall measurements. Those electrodes are fabricated on 300 nm-thick $SiO_2$ atop a heavily doped Si substrate used for back gating. We only choose the flakes landed on all four parallel electrodes with a suitable size (> 5 $\mu$m) and thickness (100-200 nm) in the first type and those landed in the gap areas in the second type. The contact resistance between the flakes and electrodes is typically ~ 1 k$\Omega$ which is stable after thermal cycling or during low-temperature transport measurements. All resistivity measurements are performed using the standard four-terminal geometry with which the contact resistance is eliminated. The transport measurements are carried out with a lock-in amplifier in an Oxford He3 cryostat with a variable temperature (T) insert which covers the *T*-range from 1.5 to 300 K and magnetic fields up to 8 T.

Since the flake sits on top of the 50 nm-thick electrodes, it is suspended above the $SiO_2$ (Fig. 1b). The atomic force microscopy (AFM) data shown in Figs. 1c and 1d indicate that the flake surface is flat. The two line scans, one over a gap between two electrodes and the other directly over an electrode, show a height difference of 50 nm, which confirms the suspended geometry. Upon exfoliation and wire-bonding, we put the mounted devices into the cryostat immediately for low-*T* transport measurements. Using this lithography-free method, we effectively minimize the uncontrolled electron density increase and therefore best preserve the desired properties (e.g.

carrier density and position of $E_F$) of the bulk materials. Indeed, the thin devices prepared from the optimally doped bulk materials show an insulating behavior which would not be possibly obtained otherwise. We also find that there are device-to-device fluctuations in carrier density, possibly due to variations in air exposure time[22] or in Ca concentration[7,8]. To compensate for prolonged air exposures which tends to increase electron density, we also deliberately choose slightly overdoped crystals, e.g. x=0.015, as our starting material (Hall measurements on the bulk crystal yield a hole density ~2 x $10^{18}$ cm$^{-3}$).

Fig. 2a shows the two-dimensional (2D) resistivity or sheet resistances $\rho_{2D}$ measured by the four-terminal method as a function of $T$ for three representative devices prepared from one crystal, among which devices $S_2$ and $S_3$ are insulating while $S_1$ is p-type metallic. Both $S_2$ and $S_3$ show a resistivity upturn below room temperature, similar to that in insulating bulk materials[7,8].

It is known that the e-beam exposure causes the electron density to increase (e.g. in HgTe)[23]. Compared with other processes in nanofabrication, however, it is relatively easier to control the dosage and energy of the e-beam. Therefore, we choose to use EBI to systematically tune the carrier density in our devices. We start with the p-type metallic device, i.e. $S_1$(M), and then introduce additional electrons step-by-step by EBI with controlled dosages and energies using a Leo SUPRA 55 scanning electron microscope (SEM) system. Under each exposure, an e-beam of moderate intensity with a fixed acceleration voltage $V_a$ and current density raster scans over the sample. $V_a$ from 4 to 16 kV has been used and the exposure time is accurately controlled so that the dosage of electrons can be determined (1.7x$10^{14}$-6.0x$10^{14}$ cm$^{-2}$). 20 kV or higher $V_a$ are found to cause the carrier mobility to decrease, indicating possible permanent material degradation. To avoid this, we limit the maximum $V_a$ to 16 kV. For alignment, only the substrate region away from the active area of the samples is exposed to the e-beam prior to the controlled low-energy EBI. During EBI, all electrodes are shorted and grounded. Fig. 2b displays the resistivity data taken immediately after each EBI. The pristine state $d_0$ of $S_1$(M) is metallic with a residual $\rho_{2D}$ of ~ 110 $\Omega/\square$ at 1.5 K. The $E_F$ initially resides in the bulk valence band, which is confirmed by the gate voltage ($V_g$) dependence as will be discussed later. Upon receiving EBI, a $\rho_{2D}$ upturn emerges at low $T$ (e.g. in $d_1$). The higher the electron dose ($d_2$, $d_3$, $d_4$, etc.), the larger the increase in the low-$T$ resistivity and consequently the onset of the resistance upturn is at a higher $T$. Qualitatively, the controlled EBI causes the metallic to insulating transition.

At sufficiently low $T$, the $\rho_{2D}$ upturn is taken over by saturation. The saturation resistivity value increases by nearly an order of magnitude, from 110 to 860 $\Omega/\square$, as the e-beam dosage progressively increases. However, the high-$T$ $\rho_{2D}$ is not at all affected by EBI, suggesting that EBI does not introduce noticeable additional scattering that would degrade the carrier mobility. The emergence of the low-$T$ insulating state could have two possible origins. One is due to the carrier localization related to the defects introduced by EBI, but it is not supported by the high-$T$ resistivity data. The other is due to the carrier density decrease which results in an upshift in $E_F$. If $E_F$ is already in the band gap but is ~ $k_BT$ away from the valence band edge, the number of thermally excited carriers will be reduced at low $T$ and the resistivity will increase. To further confirm this picture, we monitor the carrier density change using separate Hall devices prepared in the second geometry (Fig. 1a). For p-type devices, the measured hole density decreases upon EBI, accompanied by an increase in resistivity. In contrast, for n-type devices, the measured electron density increases but the resistivity decreases. Both confirm that EBI generates excess electrons which shift $E_F$ upwards. The EBI induced hole density decrease could arise from a couple of possible scenarios: trapped electrons in the material and displacement of Se atoms from their lattice sites. More detailed investigation is underway to pinpoint the actual mechanism.

At T= 0 K, the residual conductivity $\sigma_0$ has two major sources: the surface states and other extended impurity states in the bulk band gap, represented by $\sigma_0 = \sigma_s + \sigma_{gap}$, where $\sigma_s$ referring to the former and $\sigma_{gap}$ to the latter. It is difficult to accurately model the resistivity upturn at the intermediate *T*. Instead, we focus on the low-*T* data. Here we make the following crude approximation. We neglect the hopping conductivity contribution and the *T*-dependence of $\sigma_s$. Since $E_F$ is still close to the valence band edge, the *T*-dependence of the total conductivity $\sigma(T)$ is only attributed to the thermally activated bulk carriers excited from the valence band, i.e.

$$\sigma(T) = A \frac{T^{3/2}}{\rho_{d_0}(T)} e^{-(E_F - E_v)/k_B T} + \sigma_0.$$ Here $\rho_{d_0}(T)$ is the experimentally measured resistivity of the pristine metallic state of the sample whose *T*-dependence is primarily from that of the carrier mobility contribution which is assumed to be the same as in other states (e.g. $d_1$, $d_2$, etc.). In addition, we also assume that the conductivity due to the bulk carriers quickly dominates as *T* rises, for the 3D density-of-states is much larger than that of the surface states. We fit the low-*T* conductivity (Fig. 2c) and obtain the position of $E_F$ relative to the valence band edge, i.e. $E_F - E_v$, which ranges from 0.89 to 8.20 meV from $d_1$ to $d_4$. The low-*T* activated behavior is re-plotted in the inset of Fig. 2c with different slopes yielding $E_F$ positions. In the pristine state, the sample is metallic and $E_F$ is located in the valence band, hence, we do not know the exact position of $E_F$. As soon as $E_F$ leaves the valence band and move into the band gap, the activated conductivity allows us to determine the position of $E_F$. Obviously, the higher the EBI dosage, the higher the $E_F$ position in the band gap.

We perform further $E_F$ tuning by applying a gate voltage $V_g$. For TI, if we start from the valence band, the density of surface states is finite but steadily decreases as $E_F$ approaches the Dirac point in the band gap. Fig. 3 shows the effect of $V_g$ in three different insulating states of sample $S_1$: $d_1$, $d_2$, and $d_4$. In each state, the *T*-dependence of $\rho_{2D}$ for $V_g=0$ and +60 V is displayed for comparison. For a fixed $V_g$, the higher the EBI dosage is, the stronger the activated behavior becomes, and the larger the residual resistivity is. For a fixed EBI condition, the larger the $V_g$ is, the stronger the activated behavior becomes, and the larger the residual resistivity is. It is clear that both EBI and the positive $V_g$ deplete the holes and push $E_F$ upwards, consequently, drives the sample more insulating, i.e. with a larger $E_F$-$E_v$. Furthermore, the residual resistivity increases for high EBI dosages or positive $V_g$, which is resulted from the decreased surface density-of-states on the Dirac cone as $E_F$ approaches the Dirac point. Fig. 3 also reconfirms that EBI merely shifts $E_F$ just as $V_g$ does but does not alter the material properties permanently.

At the lowest $T$ (1.5 K), the thermally activated carriers are exponentially suppressed. A continuously varying $V_g$ allows $E_F$ to sweep through the extended states in the band gap. The $V_g$-dependence of $\rho_{2D}$ is shown for all EBI doses, i.e. from $d_0$ to $d_4$ in Fig. 4a. Clearly, the pristine metallic state $d_0$ has small positive $V_g$-dependence, i.e. increasing with higher $V_g$, which indicates p-type carriers. Upon EBI, $E_F$ enters the band gap, and $\rho_{2D}$ starts to rise as $V_g$ sweeps from -60 to +60 V. This occurs since large $V_g$ further depletes holes, which results in a large relatively change in hole density. With more EBI, the gate response becomes stronger. The largest $\rho_{2D}$ modulation occurs in $d_4$, i.e. from 165 Ω/□ ($V_g$ = -60 V) to 910 Ω/□ ($V_g$ = +80 V). Fig. 4b shows the corresponding conductivity $\sigma_{2D}$ as a function of $V_g$. The inset is a zoom-in $\sigma_{2D}$ plot to show a minimum of $\sim 28.5 \frac{e^2}{h}$. Note that this minimum conductivity comes from both top and bottom surfaces of the Bi$_2$Se$_3$. Therefore, each surface only contributes to about a half of this value, which agrees with some reported measurement results in Bi$_2$Se$_3$ thin flake samples.[18,21] The minimum conductivity of each surface is still much greater than $\frac{e^2}{h}$, the order of the expected minimum if only 2D massless Dirac fermions are present as in graphene.[24,25] The excess conductivity may be due to the extended impurity states in the band gap. We calculate the field-effect mobility $\mu_g$ from the slope of the $\sigma_0$ vs. $V_g$ curves using $\sigma_0 = e\mu_g n_{2D} = e\mu_g \frac{C_g}{e} V_g$, $C_g$ being the capacitance per unit area. In our device geometry, the flake is supported by 50 nm-thick Au electrodes on a 300 nm thick $SiO_2$ dielectric layer (Fig. 1), which yields $C_g$ ~70 aF/μm². Here the mobility $\mu_g$ is obtained using $\mu_g = \frac{\partial \sigma_0}{e \partial n_{2D}}$ in the region away from the conductivity minimum, i.e. where the transport is dominated by single type of carriers. As the Dirac point is approached, the conductivity vs. carrier density curve turns non-linear, but we evaluate the effective mobility in the linear region. In the other device geometry (Hall geometry), when away from the Dirac point (i.e. the conductivity minimum), we always find that the measured Hall resistivity stays linear in magnetic field, which indicates a single carrier type behavior. However, in the vicinity of the Dirac point, the Hall resistivity becomes non-linear in magnetic field, a behavior of more than one type of carriers. In this region, it is quite difficult to extract the carrier density and carrier mobility for both types. To avoid this complication, we stay in the region where the transport of single type of carriers dominates. For example, from the slope of the straight line in Fig. 4b, $\mu_g$ is found to be ~9,000 cm²/Vs, which represents the mobility of holes away from the Dirac point.

We adopt the same procedures and obtain the field-effect mobility for other EBI states. In Fig. 4c, we plot $\mu_g$ as a function of $\rho_{2D}$ evaluated for different EBI doses. In the pristine state, $\mu_g$ is as low as 800 cm²/Vs, which is the mobility of the bulk carriers in $Bi_{2-x}Ca_xSe_3$ nanodevice, which is somewhat lower than that of our p-type bulk crystals (~1200 cm²/Vs) measured separately (similar effect of exfoliation is also seen in other devices[16]). As the device turns insulating by electrostatic gating, normally we expect $\mu_g$ to drop sharply due to the impurities in the gap. On the contrary, the observed $\mu_g$ increases precipitously by more than a factor of 10, to ~ 9,000 cm²/Vs. This sharp rise in $\mu_g$ reveals that there must exist high mobility states in the band gap which are distinctly different from ordinary gap states. Here we tentatively assign those high mobility states to the surface states of 3D TI. In 3D TI, the surface carriers are massless Dirac fermions. In the absence of spin-flip mechanism, backscattering is suppressed though not strictly forbidden as in 2D TI, which should give rise to high carrier mobility. As we tune the Fermi level from the valence band to the band gap, the bulk states are replaced by the surface states, which is accompanied by a rapid decrease in conductivity but a sharp rise in mobility. We note that the surface state mobility in $Bi_2Se_3$ is already larger than the mobility of most $SiO_2$-supported graphene devices. We should point out that our devices are prepared by the lithography-free process, and the cleaner and suspended surfaces may partly contribute to the high surface carrier mobility, for the surface states are not completely immune to scatterings. By removing extrinsic disorders, we have reduced the overall scattering rate so that the intrinsic properties are better unveiled.

In summary, we have clearly demonstrated dramatic enhancement (> a factor of 10) of carrier mobility in $Bi_2Se_3$ nanodevices as we systematically control the position of $E_F$ from the valence band to the band gap via post-fabrication EBI and electrostatic gating. The mobility enhancement reveals the existence of some special high-mobility states in the band gap of $Bi_2Se_3$. We attribute those high-mobility states to the surface states of 3D TI in which the backscattering is suppressed.

Acknowledgments: we thank Chandra Varma, Allen Mills, Ward Beyermann, Harry Tom, Tao Lin, Hamad Mohamed Alyahyaei, and Dong Gui for very useful discussions and assistance. This work was supported by the US Department of Energy, Office of Basic Energy Science, Division of Materials Sciences and Engineering under Award #DE-FG02-07ER46351 (PW, ZYW and JS), and by DMEA/CNN H94003-10-2-1004 (XFL and JS).

**Figure Captions:**

Fig 1 (Color). (a) False-colored SEM image of a $Bi_2Se_3$ device taken after all measurements are completed. The exfoliated flake sits on top of four parallel electrodes. Two types of device geometries are also shown. (b) Illustration of a suspended $Bi_2Se_3$ device on top of 50 nm-thick electrodes. (c) AFM line scans over the flake surface above a gap and above an electrode. (d) AFM image of the $Bi_2Se_3$ device in (a).

Fig 2 (Color online). (a) *T*-dependence of $\rho_{2D}$ of three different samples $S_1$, $S_2$ and $S_3$ exfoliated from the same bulk crystal. $S_2$ and $S_3$ show an insulating behavior below a certain temperature. For $S_1$, both the metallic and insulating states are shown. (b) *T*-dependence of $\rho_{2D}$ of sample $S_1$ under different EBI conditions at $V_g$= +60 V. (c) Device conductivity $\sigma_{2D}$ as a function of *T* for different EBI conditions of sample $S_1$. Solid lines are the fits as described in the text. The inset shows the linear dependence of $\ln[(\sigma-\sigma_0)\rho_{d_0}T^{-3/2}]$ on 1/T, a thermally activated behavior.

Fig 3 (Color online). Effect of $V_g$ on the *T*-dependence of $\rho_{2D}$ of sample $S_1$ for the exposure doses of $d_1$, $d_2$ and $d_4$. Smaller $V_g$ negates the effect of higher dose EBI.

Fig 4 (Color online). (a) $V_g$-dependence of $\rho_{2D}$ for different EBI conditions measured at 1.5 K of sample $S_1$. As the dosage increases, $\rho_{2D}$ modulation becomes larger. (b) $V_g$-dependence of $\sigma_{2D}$ for $d_4$ at 1.5 K. The linear fit yields a field-effect mobility of ~9,000 cm$^2$/Vs. The inset magnifies the $V_g$ range from 30 to 140 V. (c) Field effect mobility plotted as a function of $\rho_{2D}$ calculated for several different states of the device at 1.5 K.

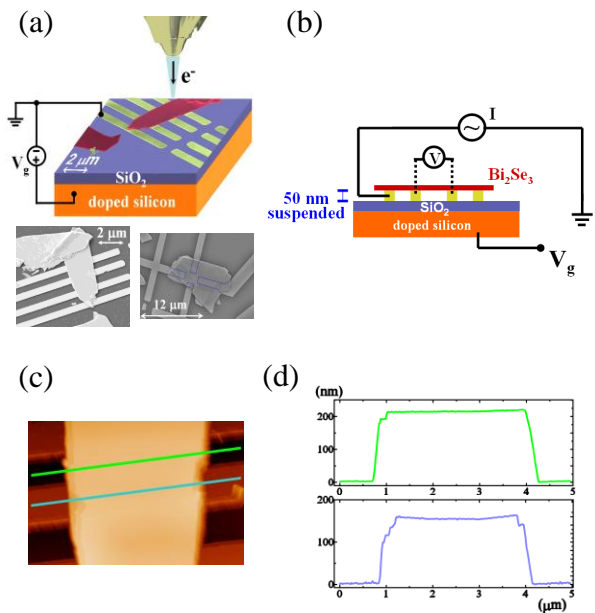

**Fig. 1**

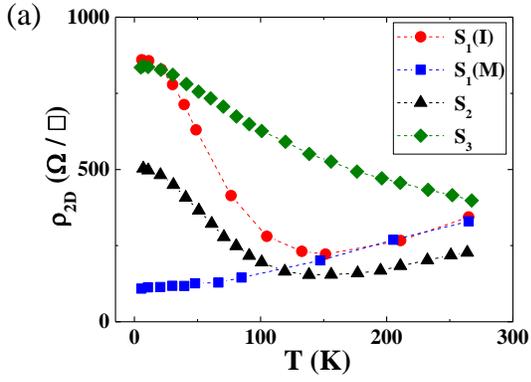

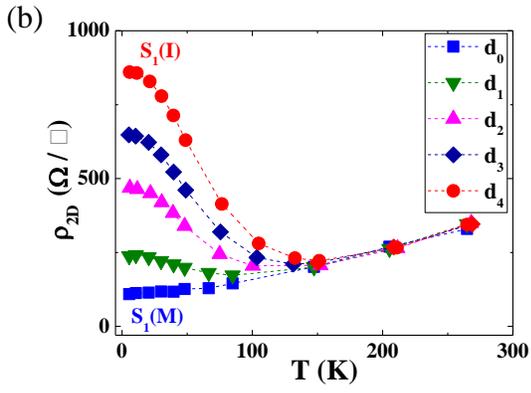

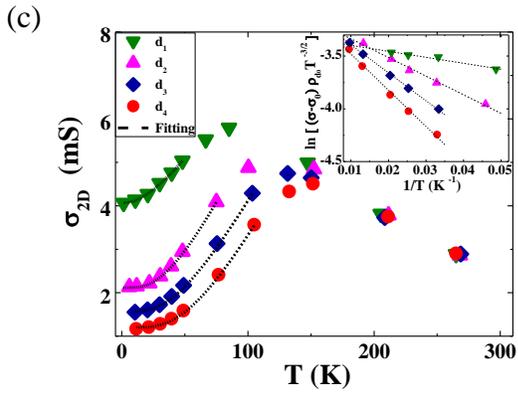

**Fig. 2**

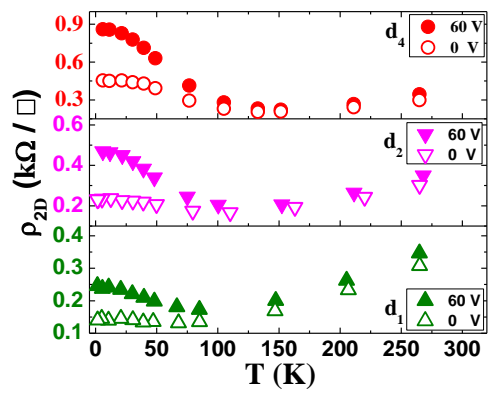

**Fig. 3**

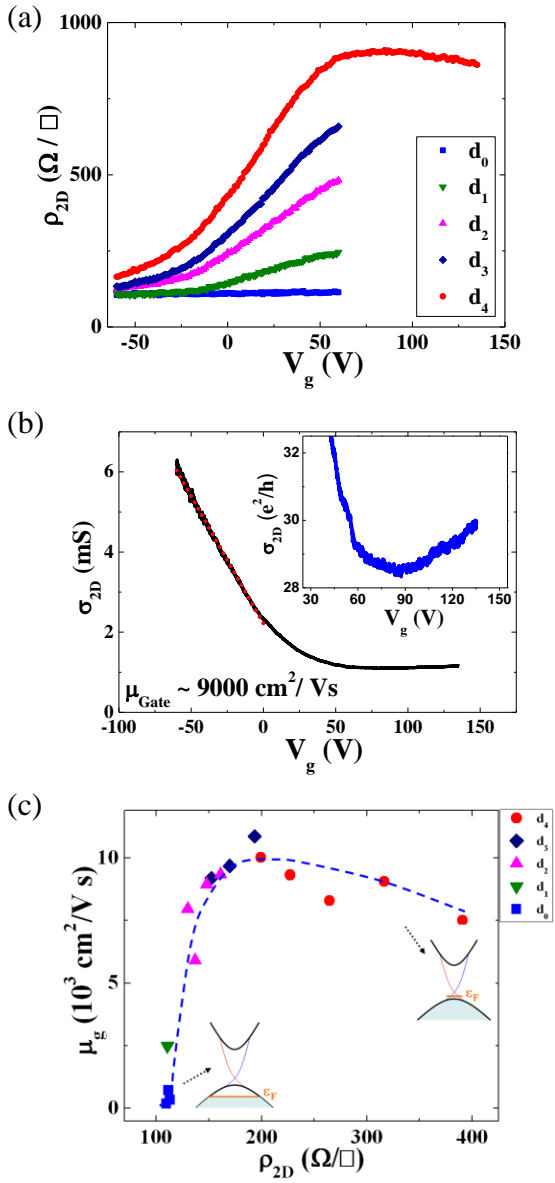

**Fig. 4**